\documentclass[12pt,eqsecnum,preprint,flushrt]{aastex6}
\usepackage{amsmath}
\usepackage{graphicx}

\newcommand{\be}{\begin{equation}}
\newcommand{\ee}{\end{equation}}
  
\newcommand{\yr}{{ \rm yr }}

\renewcommand\footnotemark{}

\begin{document}

\title{EMISSION FROM MAGNETIZED ACCRETION DISKS AROUND YOUNG STARS}

\author {C. Tapia$^1$ and S. Lizano$^1$}

\affil{$^1$Instituto de Radioastronom{\'i}a y Astrof{\'i}sica, UNAM, Apartado
Postal 3-72, 58089 Morelia, Michoac\'an, M\'exico }
\email{c.tapia@crya.unam.mx}

\begin{abstract} 
We calculate the emission of  protoplanetary disks threaded by a poloidal magnetic field and irradiated
by the central star. {The radial structure of these disks was studied by Shu and collaborators} and the vertical structure was studied by Lizano and collaborators. 
We consider disks around low mass protostars, 
T Tauri stars, and FU Ori stars with different mass-to-flux
ratios $\lambda_{\rm sys}$. We calculate the spectral energy distribution and the antenna temperature profiles
at 1 mm and 7 mm convolved with the ALMA and VLA beams. We  find that  disks with {weaker} magnetization
 (high values of $\lambda_{\rm sys}$) emit more than disks with {stronger} magnetization
(low values of $\lambda_{\rm sys}$).
This happens because the former are denser, hotter and have larger aspect ratios, receiving more irradiation from the central star.
The level of magnetization also affects the optical depth at millimeter wavelengths, being larger for disks {with high $\lambda_{\rm sys}$}. 
In general, disks around low mass protostars and T Tauri stars are optically thin at 7 mm while disks around FU Ori are optically thick.
A qualitative comparison of the emission of these magnetized disks, including heating by an external envelope,
with the observed millimeter antenna temperature profiles of
HL Tau indicates that large cm grains are required to increase the optical depth and reproduce the observed 7 mm emission at large radii.
 
\end{abstract}


\section{Introduction}
During the process of gravitational collapse and protoplanetary disk formation, 
one expects that the magnetic field from the parent dense core will be dragged into the 
disk. Nevertheless, it has been found that, if  the field remains frozen in the gas during the collapse, 
it would produce the so-called catastrophic magnetic breaking and
prevent the formation of a rotationally supported disk (RSD). To allow the formation of RSDs several processes have
been proposed like magnetic field dissipation at high densities or the reduction of the torques by 
 a misalignment between the magnetic field direction and the rotation axis (e.g., see review by 
\citealt{Lizano_2015}). The magnetic field left over from these processes  will permeate the protoplanetary disk allowing
for the operation of the magneto rotational instability (MRI) which provides a natural mechanism for the 
disk viscosity \citep{Balbus_Hawley_1998}.

Linearly polarized millimeter emission at disk scales has been observed toward a few protostars 
like IRAS 16293-2422B \citep{Rao_2014}, HL Tau \citep{Stephens_2014}, L1527 \citep{Segura-Cox_2015}, and NGC 1333 IRAS 4A1 \citep{Cox_2015}. If the polarized light is produced by emission from elongated dust grains aligned with respect to the magnetic field lines, the polarization vectors rotated by 90 degrees give the direction of the magnetic field lines (e.g., see review by \citealt{Lazarian_2007}). In these sources the magnetic field seems to have an important toroidal component, expected in young sources with substantial
infalling envelopes. Recently, \cite{Kataoka_2015} pointed out that polarization due to dust self-scattering 
by large grains in disks can be very important at millimeter wavelengths, when the maximum grain size is
$a_{\rm max} \sim \lambda/ 2\pi$, where $\lambda$ is the observing wavelength. In this case, the scattering mass opacity
due to large grains can be as large as the absorption opacity (see their Figure 1). To produce polarized light, an asymmetry in the distribution of the light source is also required.
The contributions of both elongated dust emission and self-scattering depend on the disk structure. 
In particular, \cite{Yang_2016(2)} calculated the contribution of both dust emission and scattering as a function of the
disk inclination to explain the observed polarization in NGC 1333 IRAS 4A1.
\citealt{Yang_2016} and  \citealt{Kataoka_2016} also argued that the millimeter polarization in HL Tau can be explained by dust scattering. 
 With the advent of ALMA many more
observations of disks will soon be available and it will be possible to disentangle the contribution of 
both mechanisms to obtain the morphology and structure of the disk magnetic field.

\cite{Shu_2007} (hereafter S07) studied models of the radial structure of accretion disks threaded by a poloidal magnetic field, dragged from the parent dense core during
the phase of gravitational collapse. The magnetized disk models of S07 consider a thin cold accretion disk with negligible mass
in force balance in the radial direction. The disk has sub-Keplerian rotation due to the magnetic tension 
of the poloidal field such that the rotation rate is 
$\Omega = f \Omega_K$,  where the sub-Keplerian factor is $f <1$ and  $\Omega_K = (G M_* / \varpi^3)^{1/2}$,
where $G$ is the gravitational constant, $M_*$ is the stellar mass, and $\varpi$ is the radial coordinate
(see their eq. [18]).  The disk evolves due to the viscosity $\nu$ that transfers angular momentum outward
and allows the matter in the inner regions to accrete, and the resistivity $\eta$ that allows the matter to cross field lines. 
Steady state models require that the dragging of field lines by the
accretion flow is balanced by the outward field diffusion. As found by \cite{Lubow_1994},  
this implies a small ratio of the resistivity to the viscosity (the inverse of the Prandtl number), $\eta/\nu \sim A$, where
$A$ is the disk aspect ratio. The disks models are characterized by a  mass-to-flux ratio $\lambda_{\rm sys} = (M_* + M_d) 2 \pi G^{1/2}/ \Phi$, where $M_d$ is the disk mass
and $\Phi$ is the magnetic flux threading the disk.  The disk models assume a power-law disk aspect ratio $A(\varpi) \propto \varpi^n$ 
and a viscosity $\nu$ given by their eq. (3). In these models, all the disk radial variables are power-laws (see their eqs. .[63]-[69], for the case $n=1/4$).

\cite{Lizano_2016} (hereafter L16) calculated the vertical structure of these magnetized disks, heated by viscous
and resistive dissipation and by the radiation from the central star. They discussed disks around 
YSOs with {strong magnetization} ($\lambda_{\rm sys}=4$) and disks with {weaker magnetization} ($\lambda_{\rm sys}=12$). The strongly 
magnetized disks are highly compressed by the magnetic pressure. In the case of the T Tauri disk, this 
large compression could be in conflict with the disk scale heights inferred from observations, suggesting that a 
significant amount of magnetic field has to be dissipated during the process of disk formation. Their Table 2 shows the values of the aspect ratio $ A_{\lambda_{\rm sys}} $, and the surface density  $\Sigma_{\lambda_{\rm sys}} $, for disk models with different mass-to-flux ratios.

In this paper we discuss the emission of the magnetized accretion disks studied by L16,
including the spectral energy distribution (SED) and the averaged antenna temperature at 1 mm and 7 mm convolved with the 
 ALMA and VLA beams, respectively.
In \S \ref{sec:MofS}, we describe the method of solution. In \S \ref{sec:Results}, we present the main results. In 
\S \ref{sec:Application} we compare the emission of the models with the observed mm antenna temperature profiles 
of the source HL Tau. In \S \ref{sec:Notcovered} we discuss processes not included in the present study. Finally,
 in \S \ref{sec:Conclu} we present the conclusions.

\section{ Method of Solution}
\label{sec:MofS}

We consider the emission of  magnetized disks around young stars  
subject to both viscous and resistive heating and irradiated by the central star, using the vertical structure models discussed by L16.  
We study disks around low mass protostars (LMP),  T Tauri stars, and FU Ori stars
 with different levels of magnetization measured by the mass-to-flux ratio 
$\lambda_{\rm sys}$. For the different star plus disk systems we obtain the 
SED and the averaged antenna temperature profiles at 1 mm and 7 mm.

Given the vertical structure of the disk and an inclination angle $\theta$ between the disk rotation axis and the line of sight (l.o.s.), 
we solve  the radiative transfer equation through the disk along rays parallel to the l.o.s. (coordinate $Z$), in a grid of points covering the area of the disk projected on the plane of the sky. 
Because the dust albedo at millimeter wavelengths is high for large grains as discussed in \S 1, the scattered light emissivity 
is included. We integrate along each ray the monochromatic intensity $I_\nu \, ({\rm erg \, s^{-1} \, cm^{-2} \, Hz^{-1} \, str^{-1}})$ 
and the monochromatic optical depth $\tau_\nu$ given by
\be
\frac{ d I_\nu ^{}}{d\tau_\nu}= -I_\nu + S_\nu, \quad {\rm and} \quad \frac{ d \tau_\nu }{dZ}= -\chi_\nu \rho,
\ee
where $S_\nu$ is the source function,
$\rho$ is the disk local density, and the total opacity is $\chi_\nu = \kappa_\nu + \sigma_\nu$, 
where $\kappa_\nu$ is the mass absorption coefficient, and $\sigma_\nu$ is the scattering coefficient.
The source function for an isotropically scattering medium is given by $S_\nu = (1 - \omega_\nu) B_\nu + \omega_\nu J_\nu$, where 
the albedo is $\omega_\nu = \sigma_\nu/ \chi_\nu$, 
$B_\nu$ is the Planck function at the local temperature, and $J_\nu$ is the mean intensity \citep{Mihalas_1978}. 
Following \cite{Dalessio_2001},  $J_\nu$ is approximated by the mean intensity of a vertically isothermal slab found by \cite{Miyake_1993}, given by their eq. (28). The effect of including scattering is to increase the optical depth, and, for a
given total disk opacity,  the flux from an optically thick region is reduced with respect to a purely absorbing medium, as shown in their Figure 22. 

We assume a dust  composition given by a mixture of silicates, organics, and water ice with a mass fractional abundance with respect to 
 the gas  $\zeta_{sil} = 3.4 \times 10^{-3}$, 
$\zeta_{org} = 4.1 \times 10^{-3}$, and $\zeta_{ice} = 5.6 \times 10^{-3}$,
with bulk densities $\rho_{sil} = 3.3 \, {\rm g \, cm}^{-3}$, $\rho_{org} = 1.5 \, {\rm g \, cm}^{-3}$,
and $\rho_{ice} = 0.92 \, {\rm g \, cm}^{-3}$ 
(\citealt{Pollack_1994}). 
The dust particles have a power-law size distribution, $n(a) \propto a ^{p}$, with an exponent $p=3.5$,  a minimum 
grain size $a_{\rm min} = 0.005 \, \mu$m, and maximum grain size $a_{\rm max} = 1$ mm. We assume that the dust and gas are well mixed. The solid lines in the upper panels of Figure 1 of \cite{Dalessio_2001} show 
the mass absorption coefficient $\kappa_\nu$ as a function of wavelength for $a_{\rm max} = 1$ mm, for a temperature 
$T=100$ K and $T=300$ K. At 300 K the NIR opacity features disappear because the water ice has sublimated. 
The upper right panel of their Figure 2 shows that the contribution
of troilite to $\kappa_\nu$, not included in our calculation, is small for $\lambda < 1$ mm.
We use the code of  \cite{Dalessio_2001} that includes gas opacity at $T > 1,400$ K. The opacity sources are 
free-free and bound-free transitions from neutrals and ions of H, He,  H$_2$, Si, Mg, C, molecular bands
from CO, TiO, OH and H$_2$O, and scattering by H, He,  H$_2$ and electrons
(\citealt{Calvet_1991}).  

The parameters of  the LMP, the T Tauri, and the FU Ori star plus disk systems are shown in Table 1. 
The first column corresponds to the  Young Stellar Object (YSO), the second column shows the disk mass accretion rate $\dot M_d$,
the third column shows the disk mass $M_d$, the fourth column shows the radius of the central star $R_*$, and the fifth column shows the total central luminosity $L_c$, including the accretion luminosity, that irradiates the disk surface.

 \section{{Results}}
 \label{sec:Results}
 
We calculate the structure and emission of disks around YSOs with different levels of magnetization, measured by 
the mass-to-flux ratio $\lambda_{\rm sys}$. We
consider strongly magnetized disks with {a low value of} 
 $\lambda_{\rm sys}=4$ and disks {with a weaker magnetization 
 with high values of  } $\lambda_{\rm sys}= 12$ and 24. 
Following S07, we assumed a viscosity coefficient $D=1$  for the LMP and the FU Ori disks.  
For the T Tauri disks we assume two values of the viscosity coefficient $D=10^{-2.5}$ and 0.01. As discussed by S07, a small value 
of $D$ for the T Tauri disks represents inefficient disk accretion 
due to dead zones near the disk mid-plane. 
The properties of the disk models are summarized in Table 2\footnote{These values have been slightly modified compared to those shown in Table 2 of L16 due to a new iteration scheme of the vertical structure models.}. 
The first column corresponds to the YSO and each value of the mass-to-flux ratio 
$\lambda_{\rm sys} = 4, 12$ and 24; the second column shows the
sub-Keplerian  factor $f $; the third column shows the disk aspect radio $A_{\lambda_{\rm sys}}$ at 100 AU;
 the fourth column shows the column density $\Sigma_{\lambda_{\rm sys}}$ 
at 100 AU; {the fifth column shows the vertical component of the magnetic field $B_{\lambda_{\rm sys}}$ at 100 AU}; the sixth column shows 
the disk radius $R_{\rm d}$; and the seventh column shows the ratio of the 
thermal to magnetic pressure at 1 AU, the plasma $\beta_1$
\footnote{The disk properties are obtained solving the
equations of the radial structure (S07) and the vertical structure (L16). In particular, 
the aspect ratio $A_{\lambda_{\rm sys}}$ is not a free parameter. 
It is obtained from the thermal balance where one takes
into account viscous and resistive heating and irradiation by the central star.
Once $A_{\lambda_{\rm sys}}$ is calculated, one can obtain the corresponding disk surface density and 
magnetic field according to eqs. (63) and (64) of S07. This is an iterative procedure that continues 
until $A_{\lambda_{\rm sys}}$ converges  and one obtains the full vertical structure.}.
The radial profiles of the aspect ratio, surface density, and vertical component of the magnetic field are
given by $A(\varpi) = A_{\lambda_{\rm sys}} (\varpi/100 \, {\rm AU})^{1/4}$,
$\Sigma(\varpi)  = \Sigma_{\lambda_{\rm sys}} (\varpi/100 \, {\rm AU})^{-3/4}$, and {$B_z(\varpi) = B_{\lambda_{\rm sys}} (\varpi/100 \, {\rm AU})^{-11/8}$. 
The radial component of the magnetic field at the disk surface is $B_\varpi^+ = 1.742 B_z$ (see Table 1 of S07).}
The plasma $\beta$ varies slowly with radius
$\beta(\varpi) = \beta_1 (\varpi/1 \, {\rm AU})^{1/4} $  for {disks with high $\lambda_{\rm sys}$ } where
 the departure from Keplerian rotation is small, and is constant for disks {with low $\lambda_{\rm sys}$}
(see eqs. (47) and (48) of S07).
 
 Figure \ref{fig:LMPTprof} shows the radial and vertical temperature profiles of the LMP disks  for different values of $\lambda_{\rm sys}$.
 The upper panels show the radial temperature profiles 
of  the mass weighted temperature $<T> = 2 \int_0^{\Sigma(\varpi)/2} T d\Sigma/ \Sigma(\varpi) $;
 the mid-plane temperature $T_c$,  and $T_{z_{90}}$, the temperature
 at the location of the mass surface $z_{90}$\footnote{ This surface contains 90\% of the disk mass (see discussion in 
  \S 5 of L16).}. These panels also show  
  the irradiation temperature $T_{\rm irr}$ at the  surface $z_{irr}$,
 where the irradiation from the central star is absorbed. 
The lower panels show the  vertical temperature profiles at different radii.
The star symbol corresponds to  the location of the  irradiation surface $z_{irr}$, and the diamond symbol corresponds to the location of the mass surface   $z_{90}$. The vertical temperature profiles show a temperature inversion close to the disk surface due to the external heating of the disk surface (see e.g., Figure 4 of \citealt{DAlessio_1998}).  The mid-plane temperature increases with $\lambda_{\rm sys}$ due to the increase of the  surface density: for the same disk mass $M_d$,
 disks {with weaker magnetization (high $\lambda_{\rm sys}$)} are more compact than disks {with stronger magnetization 
 (low $\lambda_{\rm sys}$)}.  

The upper panels of Figure \ref{fig:LMPSED} show the different surfaces of the LMP disks: the disk surface $z_{\infty}$, the irradiation surface $z_{irr}$, and the mass surface $z_{90}$. The lower panels show the SEDs of the star plus disk systems at different inclination angles $\theta$. 
 For a large inclination angle  $\theta = 80^\circ$, the emission from the star is occulted by the disk.
 A silicate feature at $\sim 10 \mu$m can be observed in the $\lambda_{\rm sys} = 4$ SED at $\theta = 80^\circ$. 

 Figure \ref{fig:TTTprofD2.5} shows the radial and vertical temperature profiles of T Tauri disks with viscosity coefficient $D=10^{-2.5}$
 for different values of $\lambda_{\rm sys}$.
 Figure \ref{fig:TTSEDD2.5} shows the different disk surfaces and 
   the SEDs at different inclination angles. 
 Because the disk radii are small (see Table 2),  
 we decided to explore also the T Tauri model with a viscosity coefficient $D=0.01$.  This coefficient determines the magnitude of the viscosity 
 (see eq. [42] of S07). The radial structure changes with $D$ as shown in  eqs. (63) - (69) of S07.  For $D=0.01$
 the surface density is smaller and,  for the same disk mass,  $R_d$ is larger than the models with $D=10^{-2.5}$
  (see Table 2). 
 Figures \ref{fig:TTTprofD2} and \ref{fig:TTSEDD2} show the temperature profiles, disk surfaces, and SEDs of these models.
 Because the disks are geometrically thin, the emission from the central star always contributes to the SED.
 The effect of magnetic compression is evident in the disks with $\lambda_{\rm sys} = 4$, where the height of the
 $z_\infty$, $z_{\rm irr}$, and $z_{90}$ surfaces  are lower than 
 than the models with higher values of $\lambda_{\rm sys}$.
 
 Figure \ref{fig:FUOriTprof} shows the radial and vertical temperature profiles of FU Ori disks  with different $\lambda_{\rm sys}$.
 Figure \ref{fig:FUOriSED} shows the different surfaces and SEDs. 
 The disks sizes are very small (3 - 16 AU) and correspond to the region in the disks 
 where the FU Ori outburst is expected to occur. 
 The  magnetic compression of the $\lambda_{\rm sys} = 4$ disk is evident in the top panels of
Figure \ref{fig:FUOriSED}. In this case, the disk is geometrically thin and the emission from the star contributes to the SED even at
large inclination angles. In contrast, in the disks with $\lambda_{\rm sys} = 12$ and  24,  
the star is occulted by the disk for large inclination angles and does not appear in the SED.

Radial profiles of the averaged antenna temperature $T_B$ and optical depth $\tau_\lambda$  are shown in Figures {\ref{fig:LMPprofiles} - {\ref{fig:FUOriprofiles}.
The antenna temperature is given by $T_B \equiv \lambda^2 I_\nu / 2 k_B$, where $k_B$ is the Boltzmann constant.
The antenna temperature is averaged over ellipsoidal annuli with an eccentricity given by  
$e = \sin \theta$. The disks shown have  $\theta= 60^\circ$.
The 1 mm and 7 mm profiles are shown in the upper and lower panels, respectively.
The dashed lines show the profiles convolved with the ALMA beam at 
1 mm, $\theta_{\rm ALMA} = 0.034 \arcsec$, and the VLA beam at 7 mm, $\theta_{\rm 7 \, mm} = 0.043 \arcsec$.
In general,  the convolved antenna temperature profiles decrease 
with respect to the model profiles, but are within the sensitivity levels of these facilities. 
The dotted lines correspond to the optical depth, shown in the left axis in each panel\footnote{The integration of the optical depth is
stopped at 25 for numerical convenience.}.

The averaged antenna temperature profiles  of the LMP and T Tauri disks show that {disks with weaker magnetization (high $\lambda_{\rm sys}$)} emit more than  {disks with stronger magnetization (low $\lambda_{\rm sys}$)}: 
they have larger values of the antenna temperature $T_B$ 
at a given radius.  Table 3 
gives the ratio of the fluxes at 1 mm and 7 mm with
respect to the fluxes of the $\lambda_{\rm sys}=4$ disk.  At 7 mm the flux ratios are larger than 1.  
{Disks with high $\lambda_{\rm sys}$} emit more because they are denser and hotter than disks {with low $\lambda_{\rm sys}$}. 
For the same reason, the T Tauri disk with a viscosity coefficient
$D=10^{-2.5}$ has larger values of $T_B$ at each wavelength than the T Tauri disk with $D=0.01$. 
At 1 mm, the flux ratios are smaller than 1 for the LMP and the FU Ori disks. This happens because the disks are optically thick 
 and the $\lambda_{\rm sys} = 4$ disks have larger sizes than disks {with high $\lambda_{\rm sys}$}.
The optical depth profiles in these figures also show that disks {with high $\lambda_{\rm sys}$} are more optically thick  than the 
disks {with low $\lambda_{\rm sys}$} because they are denser. Also, the $\lambda_{\rm sys} = 4$ and 12  disks 
around LMP and T Tauri stars are optically thin at 7 mm,  except  in the $\sim 1 - 2$ AU central region. 
In contrast, the FU Ori disks are optically thick because they are very small and dense; thus, the antenna temperature reaches 
large values, of the order of the kinetic temperature,    $T_B\sim T \sim 1000$ K. 
Also, the 7 mm profile of the $\lambda_{\rm sys} = 4$  FU Ori disk shows a sharp decrease in
the emission at $\sim 3$ AU.  
This happens because there is a contribution from gas opacity for $T > 1,400 $ K (see \S \ref{sec:MofS}).  These large temperatures occur only 
in the mid-plane of the inner 2 AU, nevertheless,  this region is projected
in the plane of the sky up to $\sim 3$ AU due to the disk inclination angle.

The LMP and the T Tauri disks are optically thick at 1mm and, because the disk are truncated at $R_d$, 
at an inclination angle of 60$^\circ$ {the emission from the hot disk mid-plane at outer edge of the hemisphere closest to the observer
produces an increase of the averaged antenna temperature at external disk radii.} This effect is observed in the 
1 mm profiles of the $\lambda_{\rm sys} = 24$ disks in 
the upper right panels of  Figures {\ref{fig:LMPprofiles} - {\ref{fig:FUOriprofiles}. It also produces a ``bump'' around $\sim 3 \, \mu$m in the 
FU Ori SED for $\theta=60^\circ$ and $80^\circ$.
Nevertheless, this is an artifact of the assumed disk truncation at $R_d$. Instead,  
 the surface density of viscously evolving disks   is expected to have an exponential decay, 
 $\Sigma = \Sigma_d \, (\varpi/R_d)^{-\gamma} \exp(-({\varpi/R_d})^{2-\gamma})$  \citep{Lynden-Bell_1974}.
 For the magnetized models discussed in these work $\gamma= 3/4$. 
 To obtain more realistic temperature profiles at the external radii, one needs to include this exponential region beyond $R_d$ which
 would attenuate the bright disk edge.  
 Assuming  $\rho(\varpi) \sim \Sigma/ (A(\varpi) \varpi )$, 
one obtains a radial optical depth $\tau_\nu = \int \kappa_\nu \rho\,  d \varpi = I_0 \, \kappa_\nu (\Sigma_{\rm sys} /A_{\rm sys}) (R_d/100 {\rm AU})^{-2}$, where $I_0 =  \int_1^{1+ R_{\rm ext}/R_d} u^{-2} exp (-u^{5/4}) du \sim 0.12$ for $R_{\rm ext} \sim 2 R_d$.
Assuming $\kappa_{1\, mm} \sim 0.01$, the LMP and T Tauri disks with $\lambda_{\rm sys} = 24$, 
have an optical depth  $\tau_{1mm} > 1$. Thus, an external exponential region can attenuate the bright rim of the truncated disks.
Nevertheless, the modification of the radial models of S07 to include this surface density exponential decay is out of the scope of this paper.

The surface density and temperature structure of the magnetized disk models discussed here are the result of 
a mechanical and thermal equilibrium. As example, in the next section we apply these models to an 
observed disk to obtain information of its physical properties. 

\vfill\eject

\section{{Application to HL Tau}}
\label{sec:Application}
In this section we  model the emission of LMP disks with the characteristics of the well known Class I source HL Tau 
whose spectacular structure was recently observed with ALMA (Alma partnership 2014). 
This source is  located in the Taurus cloud at a distance of 140 pc. 
The disk shows  multiple rings whose origin and physical conditions have motivated many recent observational and theoretical studies (e.g.,
\citealt{Jin_2016}; \citealt{Okuzumi_2016}; \citealt{Ruge_2016}; \citealt{Takahashi_2016}; \citealt{Yen_2016}).
Figure (3) of \citet{Carrasco-Gonzalez_2016} shows the antenna temperature profiles of HL Tau at 0.87 mm, 1.3 mm, 2.9 mm, and 7 mm, 
obtained with ALMA and VLA. We make a qualitative comparison of the emission of the magnetized disk models with these temperature profiles to obtain general properties of the HL Tau disk, instead of modelling the detailed ring structure as done recently, for example, by 
\citet{Pinte_2016} with a parametrized disk structure.  

\citet{Dalessio_1997} showed that irradiation by the HL Tau envelope is needed to 
heat the disk and raise its temperature in the outer regions to reproduce the  observed fluxes at mm wavelengths.  
Thus, to compare with the observed mm profiles, we include a simple envelope heating: we assume that the disk is irradiated 
by a thermal bath with a temperature $T_e$ such that, 
 at the disk surface,  the mean intensity due to the envelope irradiation is $J_e = \sqrt{3} \sigma T_e^4 / (4 \pi) $. 
Then, $J_e$ is added to the B.C. in eq. (45) of L16, and the envelope flux $(4 \pi / \sqrt{3}) J_e e^{(-\tau_d)}$ 
is also added to the reprocessed flux in their eq. (41), where $\tau_d$ is the opacity normal to the disk mid-plane.
 
HL Tau has observational estimates of disk mass, radius, accretion
rate, and luminosity  which are a little different from the  reference LMP model discussed in
Section 3. For example, the  disk mass of HL Tau is 50 \% higher and
the disk accretion rate is  50\% lower than the LMP model. The disk radius is obtained
from eq. (65) of S07, with $A_{\rm sys}$ calculated self-consistently from the vertical structure modelling (L16). We find
that the $\lambda_{\rm sys} = 12$ models are too large, with $R_d = 466$ AU. For this reason, 
we consider  magnetized disk models with $\lambda_{\rm sys} = 24$ which have $R_d \sim $100 AU, as
observed in this source (\citealt{Kwon_2011}).

Table  4 
shows the parameters chosen for the HL Tau disk: the mass accretion rate $\dot M_d$, the disk mass $M_{d}$, 
 the luminosity of the central source $L_c$ 
that includes the stellar and the accretion luminosities, the inclination angle $\theta$, 
the mass-to-flux ratio $\lambda_{\rm sys}$, and the sub-Keplerian factor $f$. We assume a central star with mass $M_* = 1 \, M_\odot$, 
radius $R_*= 2.2 \, R_\odot$ and a
temperature $T_* = 4,000$ K. 
We discuss 6 LMP disk models (Model I - VI) that have different envelope temperatures
$T_e = 0, 50, 100 $ K and different values of the maximum grain size $a_{\rm max} = 1 \, {\rm mm},   1 $ cm. For these models,
Table 5 shows the aspect ratio $A_{\lambda_{\rm sys}}$, the mass surface density $\Sigma_{\lambda_{\rm sys}}$ at 100 AU, 
{the vertical component of the magnetic field $B_{\rm sys}$ at 100 AU, }
the disk radius $R_d$, and the plasma $\beta_1$ at 1 AU.

The upper panels of Figure \ref{fig:HLTauprofiles}  show the convolved antenna temperature profiles $T_B$ at 
 0.87 mm, 1.3 mm,  2.9 mm,   and 7 mm.
The lower panels show the corresponding convolved optical depth profiles $ \tau_{\lambda}$. 
 The left panels correspond to Model I (dotted lines), Model II (solid lines), and Model III (dashed lines)
 which have a dust distribution with $a_{\rm max} = 1$ mm.
The emission is optically thick from 0.87 mm to  2.9 mm,  thus,  Model I without envelope heating is too cold to reproduce the 
ALMA profiles. On the other hand,  Model III, which has a substantial envelope heating ($T_e = 100$ K), 
overestimates the antenna temperatures at these wavelengths. 
 The observed 7 mm profile cannot be reproduced by
 Models I - III.  Because the emission at 7 mm is optically thin, the 7 mm profiles have very low temperatures at 100 AU. 
In order to increase the opacity, we decided to explore models with a dust distribution with $a_{\rm max} = 1$ cm.
 The right panels correspond to Model IV (dotted lines), Model V (solid lines), and Model VI (dashed lines), which have a dust 
 distribution with  $a_{\rm max} = 1$ cm. The effect of the large grains is to increase the optical depth at 7 mm.
  As before, the temperature profiles of Model IV (with no envelope heating) 
 are too low at all wavelengths, while the temperature profiles of  Model VI ($T_e = 100$ K) are too high for 0.87 mm to 2.9 mm. 
 
 The observed temperature profiles of  \citet{Carrasco-Gonzalez_2016}  are best reproduced by Model V
which has a moderate envelope heating ($T_e = 50$ K). This model also reproduces the observed 7 mm
VLA profile. This happens because its opacity increased by a factor of 10 with respect to the models with $a_{\rm max} = 1$ mm.
Note that the 7mm opacity reported by Carrasco-Gonz\'alez et al. (2016) is a factor of $\sim 15$ lower than the opacity of Model V. 
Nevertheless, Carrasco-Gonz\'alez et al. (2016) assumed a dust temperature profile
and, from the simple equation of radiative transfer
$T_B = T_d (1- e^{-\tau_\nu})$, solved for the optical depth $\tau_\nu$ (see the first 
paragraph in Section 3.1 of their paper) . Since Model V reproduces the level of 
observed emission at 7 mm,  if one applies the same procedure, 
one would obtain optical depths similar to their values. Instead, what is plotted in the lower
panels of Figure  \ref{fig:HLTauprofiles}  is the physical optical depth profiles of the models convolved with the ALMA and VLA beams.

 From this qualitative study we find that it is difficult to reproduce the observed emission at 7 mm of the HL Tau disk  at large radii
 just including the  envelope heating. 
 We conclude that  one possibility is that the HL Tau disk has large grains at the external radii which can increase
 the optical depth. 
 Then, with both envelope heating and large grains, Model V can produce  the observed level of 7 mm emission at the external radii.
  
\section{{Physical processes not included in this study}}
\label{sec:Notcovered}

In the models discussed in this work we have not considered several
processes. These include dust growth, settling and radial migration which are expected to occur in protoplanetary disks (for a review
see \citealt{Williams_2011}). Also, we do not include physical processes like the formation of vortices
or spiral arms that have been observed in several sources (e.g., \citealt{van_der_Marel_2016};  \citealt{Perez_2016}).

Grain growth can be taken into account by
considering different values of $a_{\rm max}$. 
Dust settling can be included by considering 
an atmospheric layer with small grains and a mid-plane layer
with larger grains such that the dust mass missing from the atmospheric layer
is incorporated into the mid-plane layer (\citealt{Dalessio_2006}). 
The degree of settling is measured by the ratio of the dust to gas mass ratio of the small grains in the atmospheric layer 
to the total dust to gas mass ratio,
$\epsilon=\zeta_{\rm small}/\zeta_T \le 1$. The settling scale height is usually a free parameter 
although it could be established by the balance of gravitational sedimentation
and turbulent diffusion (e.g., \citealt{Dubrulle_1995}),  or  from observations as in the case of
\citet{Pinte_2016} who inferred a very thin dust disk in the case of HL Tau.
Dust radial drift has been studied by many authors (e.g., \citealt{Takeuchi_Lin_2002}; \citealt{Brauer_2008}; \citealt{Birnstiel_2010}). Their models 
show that this process should occur in very short timescales, in conflict with the observations of disks that infer
 mm and cm dust grains in the external regions of disks. Nevertheless, recent high resolution mm observations have found a radial gradient
in dust sizes in several sources (e.g., \citealt{Perez_2015}; \citealt{Tazzari_2016}). Then, to include the effect of the dust radial drift
one can assume a radial variation of $a_{\rm max}$ together with a variation of $\zeta_T$ .  
The inclusion of all these processes will be the subject of a future study.

It will be interesting to model other T Tauri and FU Ori sources when high spatial resolution 
 ALMA and VLA data will be available, that can provide information about 
 the disk temperature and optical depth, as in the case of HL Tau. In particular, to study older disks, processes like
 settling and radial migration need to be included in our models.

Finally, the relevant range of values of $\lambda_{\rm sys}$ in protoplanetary disks will eventually come from observations or
 from models and numerical simulations of disk formation. 

 \section{Conclusions}
\label{sec:Conclu}

We calculate the emission of magnetized accretion disks irradiated by the central star using the vertical 
structure models of L16. We consider disks with different levels of magnetization, measured by the
mass-to-flux ratio $\lambda_{\rm sys}$. We include the SED and the averaged antenna temperature 
profiles at 1 mm and 7 mm, convolved with highest angular resolution beams of ALMA and VLA. 
 
 We find that disks with {weaker magnetization} (high values of the 
 mass-to-flux ratio $\lambda_{\rm sys}$) emit more than disks with {stronger magnetization} (low values of 
 $\lambda_{\rm sys} $). This happens because the former disks are denser and have larger aspect ratios. 
 Thus, they receive more irradiation from the central star and are hotter than the {more} strongly magnetized disks. 
The optical depth at millimeter wavelengths also varies with the level of magnetization because 
{disks with high $\lambda_{\rm sys}$}  are denser than {disks with low $\lambda_{\rm sys}$}.  Disks around LMP and T Tauri stars are optically thick at 1mm and are optically thin at 7 mm. 
Instead, the FU Ori disks are always optically thick.

We compare the emission of magnetized disk models with observed mm antenna temperature profiles 
of the disk of  HL Tau. We find that models with a dust distribution with a maximum grain size $a_{\rm max} = 1$ mm do not reproduce 
the observed 7 mm profile, even including the heating due to the envelope irradiation. 
Because the emission is optically thin, the
7 mm antenna temperature drops to very low values at large radii. One possibility is 
 the HL Tau  disk has large grains,  with $a_{\rm max} = 1$ cm, that increase the dust opacity. 
 Then, together with the envelope heating, the disk can reach the observed 7 mm emission  at the external radii.

In the near future, on expects that high angular resolution observations of magnetic fields from disks 
 around young stars will be obtained with the ALMA and VLA  interferometers. It will be very useful to compare these
 observations with the structure and emission of the magnetized disks models discussed in this work.
 This comparison can help constrain the level of magnetization in protoplanetary disks, measured by
 their mass-to-flux ratio, to understand their formation and evolution.
 
 \acknowledgments

CT and SL acknowledge support by  CONACyT 153522 and UNAM-PAPIIT 105815. They also acknowledge  valuable comments
and suggestions from an anonymous referee which helped improved this manuscript.

\begin{deluxetable}{lllll}
\tablecolumns{5}
\tablewidth{0pc}
\tablecaption{YSOs parameters }
\tablehead{
 \colhead{YSO} & \colhead{$\dot M_{d} $} &  \colhead{$M_d$}  
 &\colhead{$R_*$} & \colhead{$L_c $} \\
 & $(M_\odot \yr^{-1})$ &$(M_\odot)$ &  $(R_\odot)$ &  $(L_\odot)$
  }
\startdata
LMP & $2\times 10^{-6}$ &0.20 &  3 &  7.1 \\ 
T Tauri  & $1\times 10^{-8}$  &  0.03 &  2 & 0.93  \\
FU Ori & $2\times 10^{-4}$ &  0.02  &  7 & 230  
\enddata 
\label{table:1}
\end{deluxetable}

\vfill\eject

 \begin{deluxetable}{lllllll}
\tablecolumns{7}
\tablewidth{0pc}
\tablecaption{Models with different mass-to-flux ratios $\lambda_{\rm sys}$}
\tablehead{
 \colhead{YSO} & $f$ &  \colhead{$A_{\lambda_{\rm sys}}$} & $\Sigma_{\lambda_{\rm sys}}$ & $B_{\lambda_{\rm sys}}$ & $R_{d}$ & $\beta_1$  \\
                          &      &                                                           & $( {\rm g/cm^2})$                      & mG                                     & (AU)   &  }
\startdata
LMP &&&&&\\
$\lambda_{\rm sys} =4$         & 0.9565 & 0.156 & 5.30  & 6.93  & 457  & 4.26 \\
$ \lambda_{\rm sys}=12$      &  0.9953 & 0.284 &  27.4 & 5.26  & 124 &  28.3 \\
  $\lambda_{\rm sys}=24$     &  0.9988 & 0.372 & 84.1  & 4.60  & 50.1 & 109 \\ 
T Tauri ($D=10^{-2.5}$) &&&&&\\
$\lambda_{\rm sys} =4$        & 0.6579  & 0.0123 & 11.0 & 25.8 & 56.0  & 2.92 \\
$ \lambda_{\rm sys}=12$      &  0.9679 & 0.102  & 17.6  & 10.9 & 38.2  & 3.85 \\
 $\lambda_{\rm sys}=24$      &  0.9921 & 0.193  &  38.2 & 8.01 & 20.6  & 10.0 \\
T Tauri ($D=0.01$)  &&&&&\\
$\lambda_{\rm sys} =4$        & 0.6579 & 0.0101 & 4.25  & 16.0  & 120   & 2.81 \\
$ \lambda_{\rm sys}=12$      & 0.9679 & 0.090   &  6.33 & 6.53  & 86.9  & 3.70  \\
 $\lambda_{\rm sys}=24$       & 0.9921 & 0.163  & 14.2  & 4.89  & 45.5  & 9.12 \\
  FU Ori  &&&&&\\
  $\lambda_{\rm sys} =4$       & 0.3865 & 0.101  & 33.3 & 55.0 & 16.7 & 2.96 \\
 $ \lambda_{\rm sys}=12$      &  0.9516 & 0.502 & 148 & 38.7 & 5.61 & 4.89 \\
  $ \lambda_{\rm sys}=24$     & 0.9881 & 0.581 & 533 & 36.7  & 2.56 & 14.5
\enddata 
\tablecomments{The radial profiles of the aspect ratio, surface density, and vertical component of the magnetic field are
given by  $A(\varpi) = A_{\lambda_{\rm sys}} (\varpi/100 \, {\rm AU})^{1/4}$, 
 $\Sigma(\varpi)  = \Sigma_{\lambda_{\rm sys}} (\varpi/100 \, {\rm AU})^{-3/4}$,  and
 $B_z(\varpi) = B_{\lambda_{\rm sys}} (\varpi/100 \, {\rm AU})^{-11/8}$. }
\label{table:2}
\end{deluxetable}

\vfill\eject

 \begin{deluxetable}{lllll}
\tablecolumns{5}
\tablewidth{0pc}
\tablecaption{Flux ratios }
\tablehead{
 \colhead{YSO} & $F_{1 \rm mm}^{12} / F_{1 \rm mm}^{4} $ &  $F_{1 \rm mm}^{24} / F_{1 \rm mm}^{4} $ & $F_{7 \rm mm}^{12} / F_{7 \rm mm}^{4} $& $F_{7 \rm mm}^{24} / F_{7 \rm mm}^{4} $ \\
  & & &  }
\startdata
LMP     & 0.90 & 0.58  & 3.01 & 5.49  \\ 
T Tauri ($D=10^{-2.5}$) &  1.62 &  1.05 &  1.58&  2.49 \\
T Tauri ($D=0.01$) & 2.23  & 1.89  &  1.91   &  3.19   \\
  FU Ori  &  0.61 & 0.39  &  1.95 & 1.37
\enddata 
\tablecomments{ \\LMP disk:  $F^4_{1\rm mm} = 3.94 \times10^{-1}$ Jy  ;  $F^4_{7\rm mm} =2.11 \times 10^{-4}$ Jy . \\
T Tauri disk ($D=10^{-2.5}$):  $F^4_{1\rm mm} =  1.30\times10^{-2}$Jy ;     $F^4_{7\rm mm} =3.54 \times 10^{-5}$ Jy\\
T Tauri disk ($D=0.01$):   $F^4_{1\rm mm} =   1.68 \times10^{-2}$ ;            $F^4_{7\rm mm} =1.74 \times 10^{-5}$ Jy .\\
Fu Ori disk: $F^4_{1\rm mm} =1.75 \times 10^{-1}$ Jy ;                           $F^4_{7\rm mm} =1.16 \times 10^{-3}$ Jy .}
\label{table:3}
\end{deluxetable}

\begin{deluxetable}{rcrrrc}
\tablecolumns{6}
\tablewidth{0pc}
\tablecaption{HL Tau parameters}
\tablehead{
 \colhead{$\dot M_{d} \, ^a $}  & \colhead{$M_d \, ^b$}  
& \colhead{$L_c $}&   $\theta \, ^c$ & $\lambda_{\rm sys}$ & $f$\\
 $(M_\odot \yr^{-1})$ & $(M_\odot)$   &  $(L_\odot)$ &  (deg)  & & 
  }
\startdata
$1\times 10^{-6} $ &   0.3  &  8.6 &  47 &  24 & 0.9984   
\enddata 
\tablecomments{Values taken from: (a) \cite{Dalessio_1997}; (b) \cite{Carrasco-Gonzalez_2016}; (c)  \cite{ALMA_2015} .}
\label{table:4}
\end{deluxetable}

\begin{deluxetable}{cccccccc}
\tablecolumns{7}
\tablewidth{0pc}
\tablecaption{HL Tau models }
\tablehead{ \colhead{Model} & \colhead{$a_{\rm max} $ } & \colhead{$T_e$}   & $A_{\lambda_{\rm sys}}$ & 
$\Sigma_{\lambda_{\rm sys}}$ & $B_{\lambda_{\rm sys}}$ & $R_d$ & $\beta_1$ \\
  &   & K &  &  (${\rm g/cm^2}$) & mG & (AU)
  }
\startdata
I &  1 mm   & 0     & 0.215  & 38.6   & 5.09  &  129   & 63.4 \\ 
II &  1 mm  & 50   & 0.237  & 35.0   & 4.85  &  140   & 61.1 \\ 
III & 1  mm & 100 & 0.268  & 30.9   & 4.56  &  154   & 59.4 \\  
IV & 1 cm   & 0     & 0.195  & 42.5   & 5.35  & 119   &  54.1 \\
V & 1 cm    & 50   & 0.225  & 36.8   & 4.98  &  134  & 53.4 \\ 
VI & 1 cm   & 100 & 0.251  & 33.0   & 4.71  & 146   & 52.5  
\enddata
\label{table:5}
\tablecomments{The radial profiles of the aspect ratio, surface density, and vertical component of the magnetic field are
given by  $A(\varpi) = A_{\lambda_{\rm sys}} (\varpi/100 \, {\rm AU})^{1/4}$, 
 $\Sigma(\varpi)  = \Sigma_{\lambda_{\rm sys}} (\varpi/100 \, {\rm AU})^{-3/4}$,  and
 $B_z(\varpi) = B_{\lambda_{\rm sys}} (\varpi/100 \, {\rm AU})^{-11/8}$.}
\end{deluxetable}

\begin{figure} 
\centering
\includegraphics[angle=0,width=1.\textwidth]{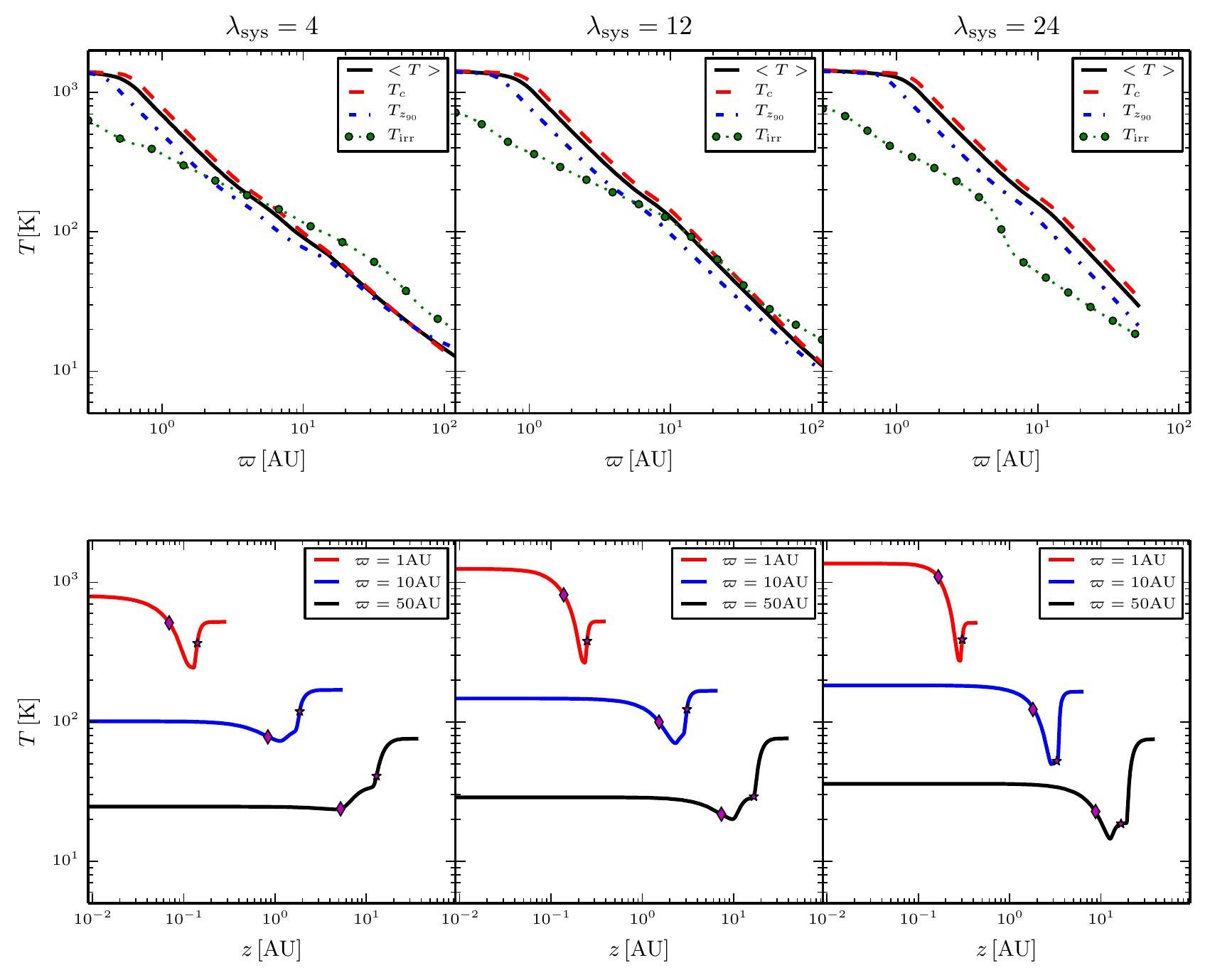}
\caption{
LMP disk models with different mass-to-flux ratios, $\lambda_{\rm sys}=4, \, 12, \, 24$, that label each column.
The upper panels show the radial temperature profiles of the disks:
 the  solid black lines correspond to the mass weighted temperature $<T>$;
 the  red dashed lines show the mid-plane temperature $T_c$;  the blue dot-dashed lines 
 show the  temperature of  mass surface $z_{90}$;  
 the green dot lines show the temperature at the irradiation surface $z_{\rm irr}$. 
The lower panels show vertical temperature structure at the radii indicated in the upper right boxes.
The star symbol corresponds to  the location of the  irradiation surface $z_{\rm irr}$ and the diamond symbol corresponds to the location of the mass surface $z_{90}$.
}
\label{fig:LMPTprof}
\end{figure}

\begin{figure} 
\centering
\includegraphics[angle=0,width=1.\textwidth]{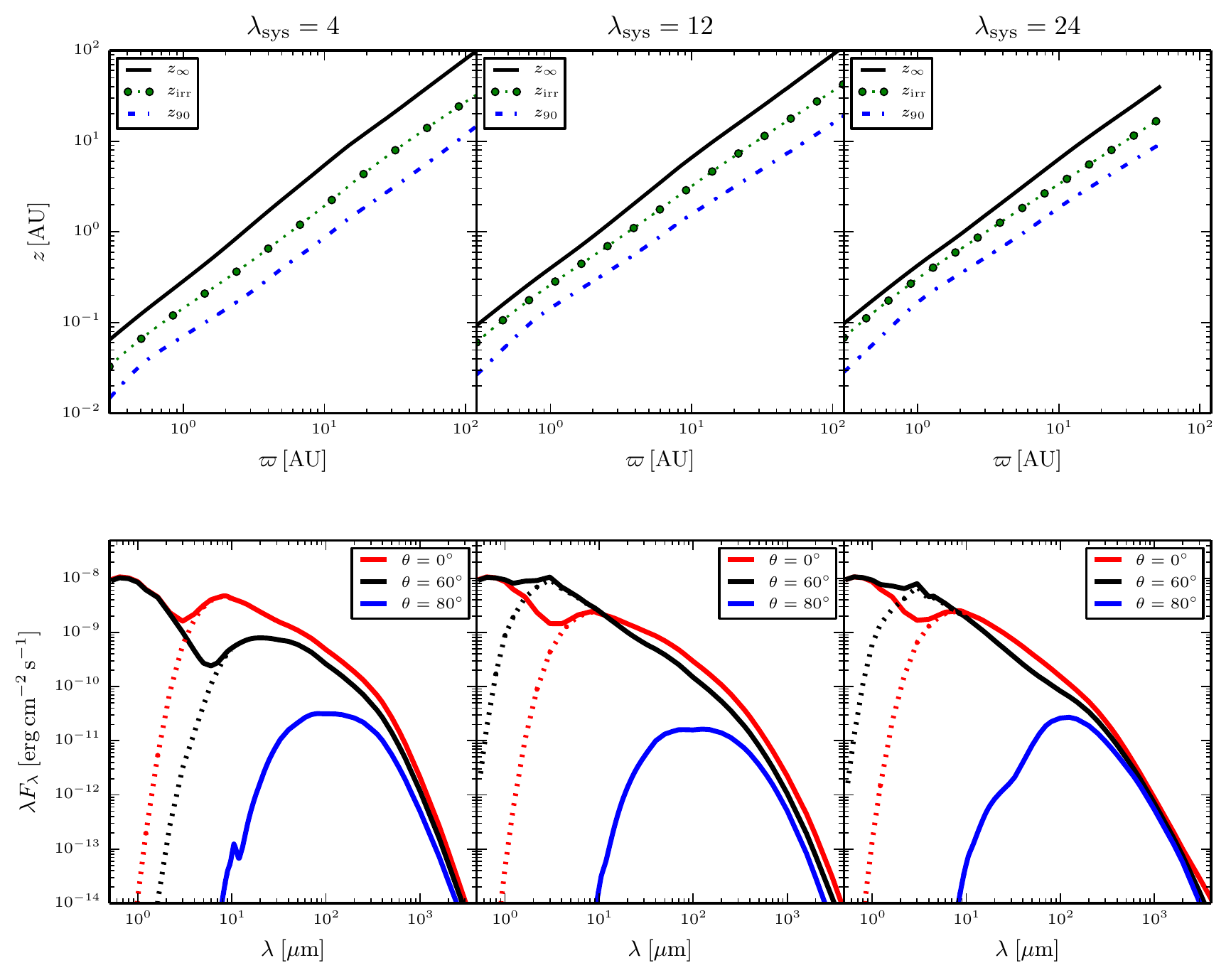}
\caption{
LMP disk models with different mass-to-flux ratios, $\lambda_{\rm sys}=4, \,12, \, 24$, that label each column.
The upper panels show the  the different surfaces:
the black solid  lines show the surface of the disk $z_{\infty}$; 
the green dot lines show  the irradiation surface $z_{\rm irr}$;
the blue dot-dashed lines 
 show the disk mass surface $z_{90}$.
 The lower panels show the spectral energy distribution (SED) of the star plus disk system at different inclination angles $\theta$ between the disk rotation axis and the l.o.s: $\theta=$ $0^\circ$, $60^\circ$, and $80^\circ$ (red, black and blue lines, respectively). The dotted lines show the disk contribution.
}
\label{fig:LMPSED}
\end{figure}

\begin{figure} 
\centering
\includegraphics[angle=0,width=1.\textwidth]{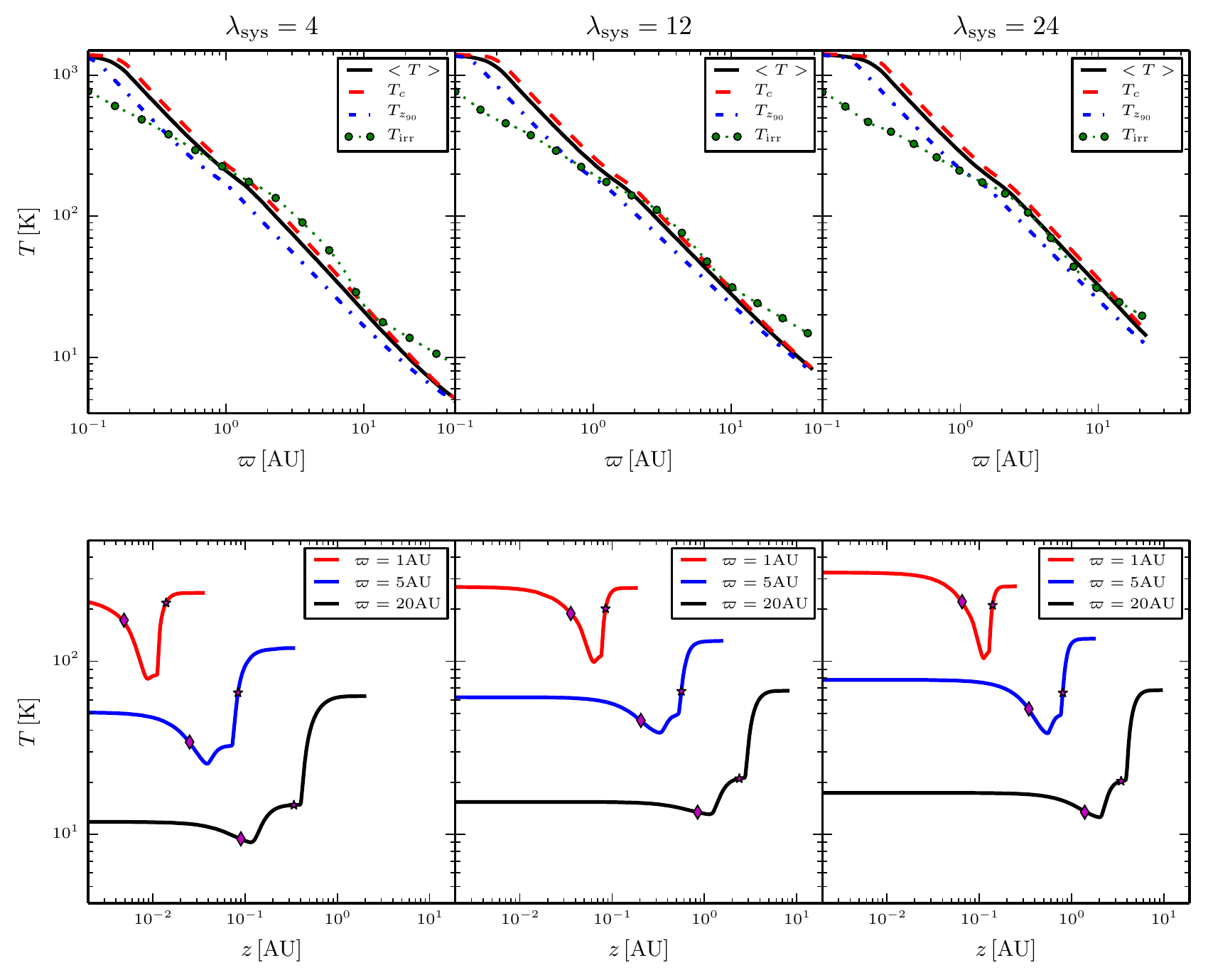}
\caption{
T Tauri disk models with viscosity coefficient $D=10^{-2.5}$ and 
 different mass-to-flux ratios, $\lambda_{\rm sys}=4, \, 12, \, 24$,  that label each column.
 The upper panels show the radial temperature profiles of the disks.
 The lower panels show vertical temperature structure at  the radii indicated in the upper right boxes.
Same description as Figure 1.
}
\label{fig:TTTprofD2.5}
\end{figure}

\begin{figure} 
\centering
\includegraphics[angle=0,width=1.\textwidth]{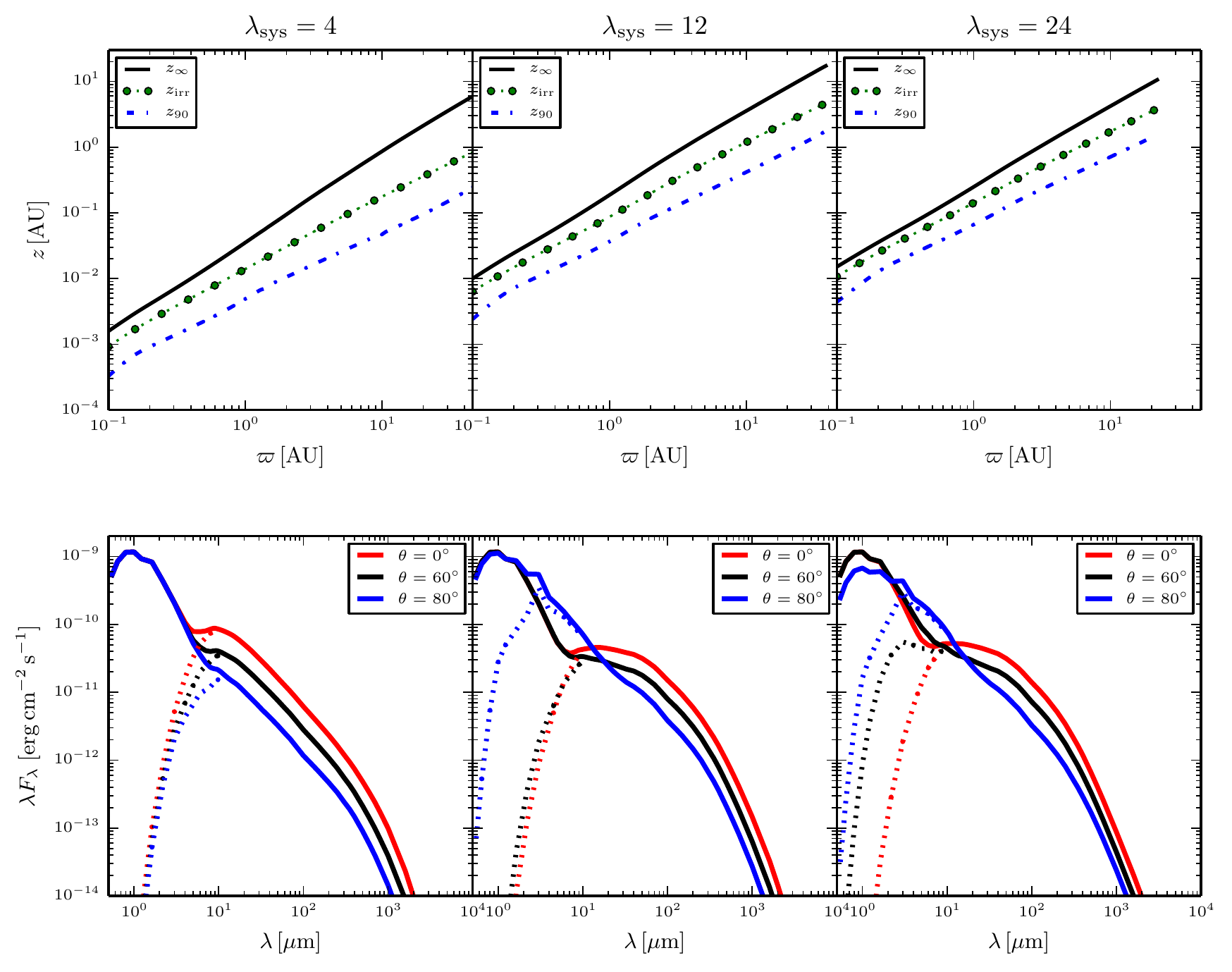}
\caption{
T Tauri disk model with a viscosity coefficient $D=10^{-2.5}$ and 
 different mass-to-flux ratios, $\lambda_{\rm sys}=4, \, 12, \, 24$,  that label each column.
The upper panels show the  the different disk surfaces.
 The lower panels show the SED of the star plus disk system at different inclination angles.
 Same description as Figure 2.}
\label{fig:TTSEDD2.5}
\end{figure}

\begin{figure} 
\centering
\includegraphics[angle=0,width=1.\textwidth]{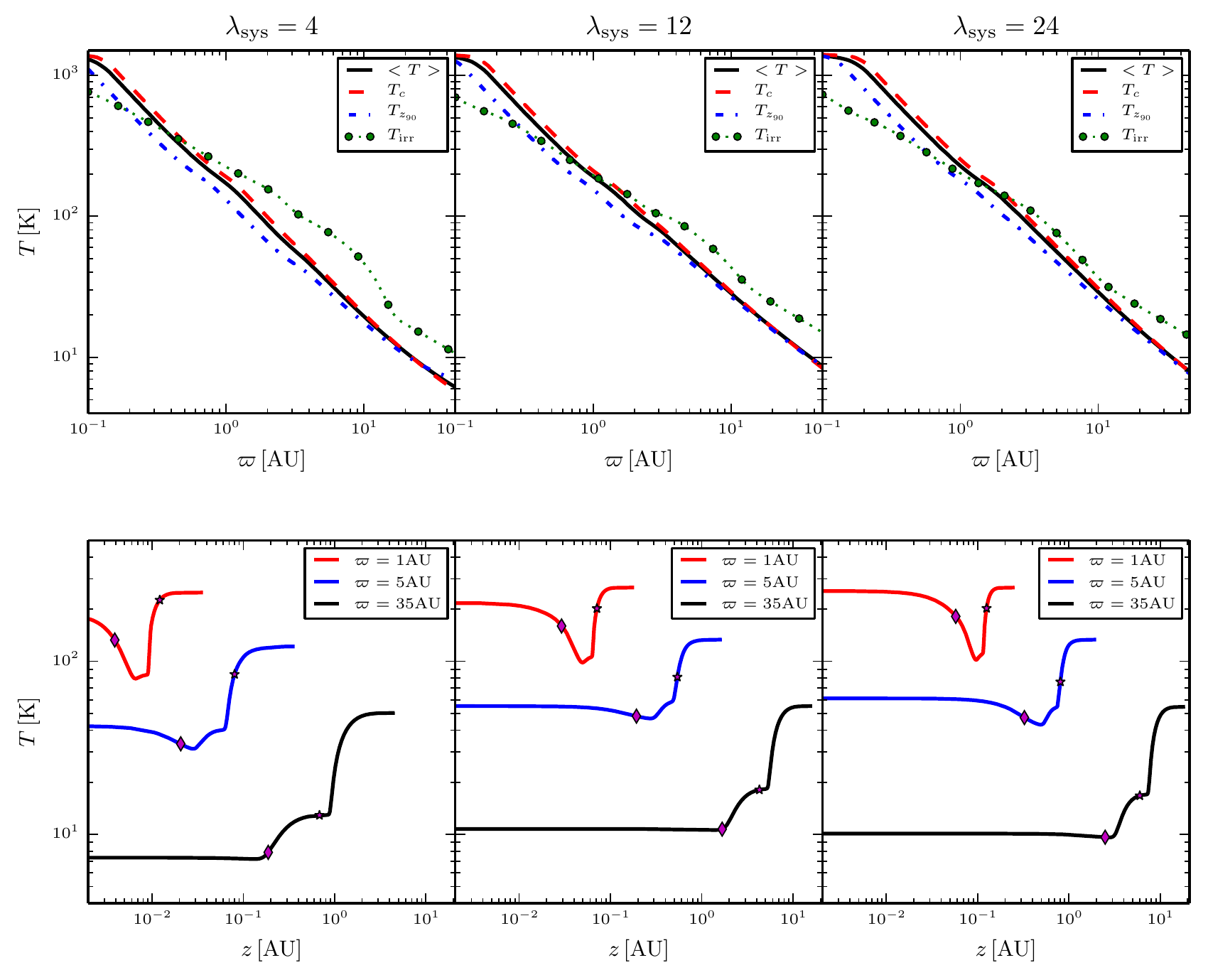}
\caption{
T Tauri disk models with viscosity coefficient $D=0.01$ and 
 different mass-to-flux ratios, $\lambda_{\rm sys}=4, \, 12, \, 24$,  that label each column.
 The upper panels show the radial temperature profiles of the disks.
 The lower panels show vertical temperature structure at  the radii indicated in the upper right boxes. 
Same description as Figure 1.
}
\label{fig:TTTprofD2}
\end{figure}

 \begin{figure} 
\centering
\includegraphics[angle=0,width=1.\textwidth]{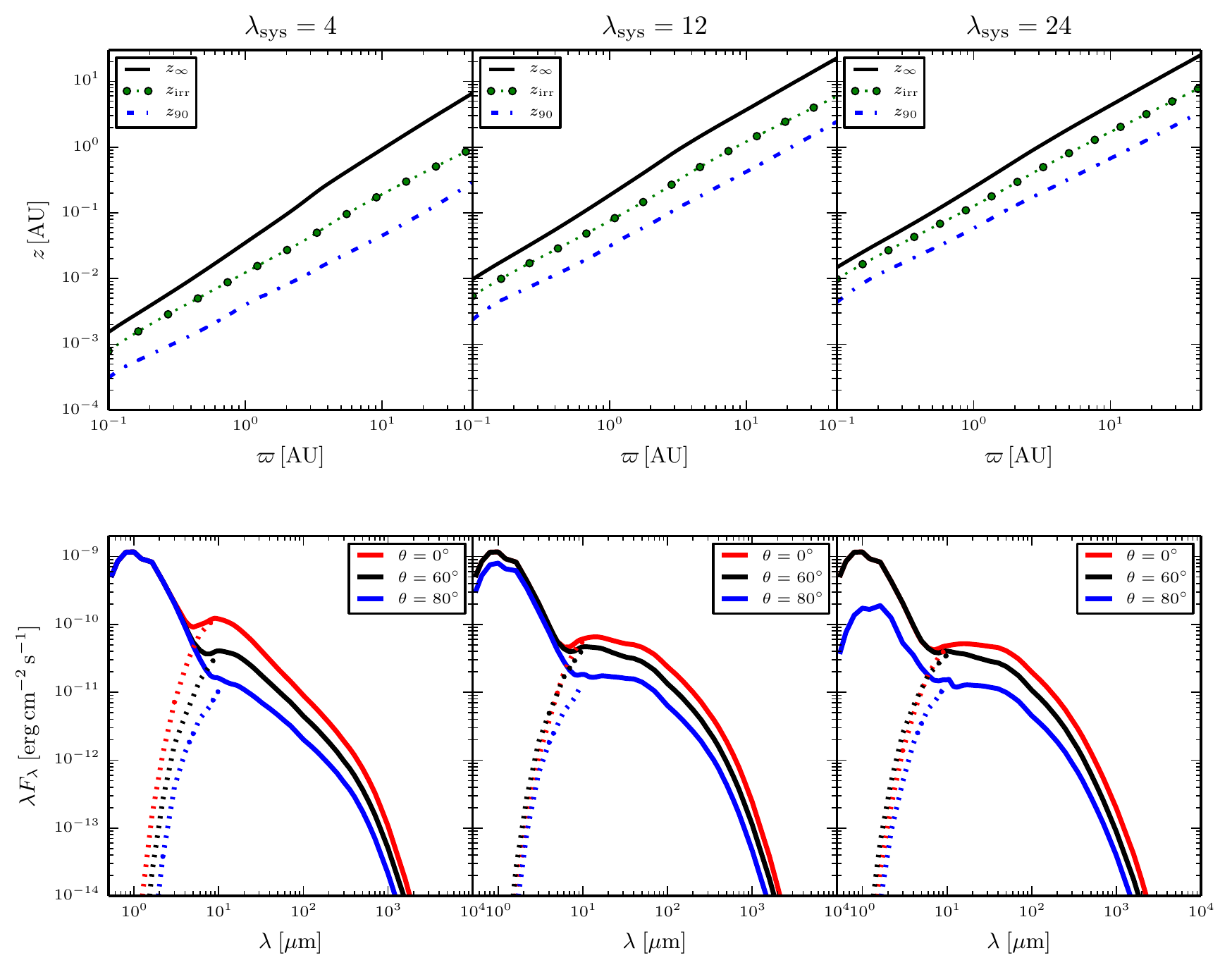}
\caption{
T Tauri disk model with a viscosity coefficient $D=0.01$ and 
 different mass-to-flux ratios, $\lambda_{\rm sys}=4, \, 12, \, 24$,  that label each column.
The upper panels show the  the different disk surfaces.
 The lower panels show the SED of the star plus disk system at different inclination angles.
 Same description as Figure 2.
}
\label{fig:TTSEDD2}
\end{figure}

\begin{figure} 
\centering
\includegraphics[angle=0,width=1.\textwidth]{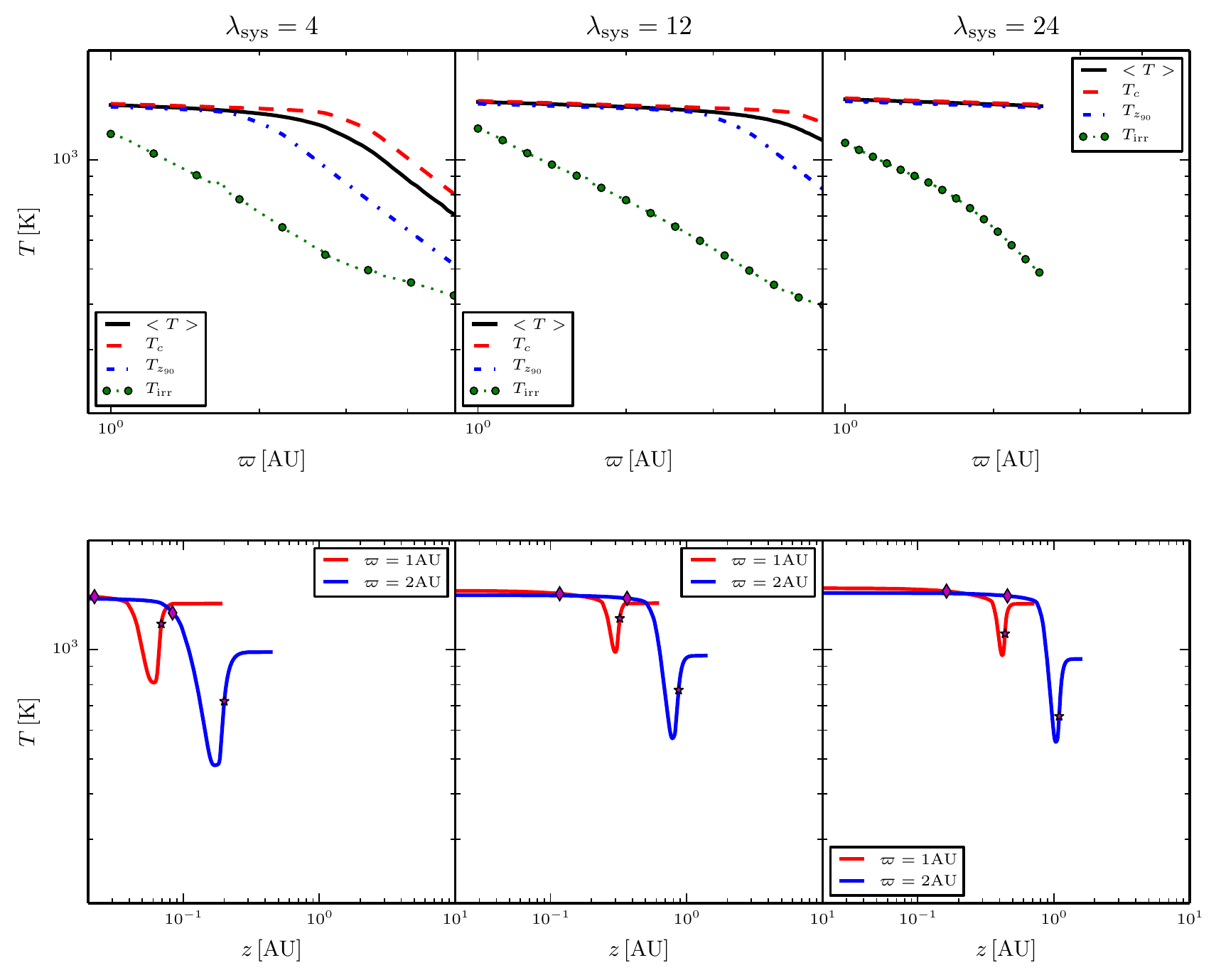}
\caption{
FU Ori disk models with different mass-to-flux ratios, $\lambda_{\rm sys}=4, \, 12, \, 24$, that label each column.
 The upper panels show the radial temperature profiles of the disks.
 The lower panels show vertical temperature structure at  the radii indicated in the upper right boxes.
Same description as Figure 1.
}
\label{fig:FUOriTprof}
\end{figure}

\begin{figure} 
\centering
\includegraphics[angle=0,width=1.\textwidth]{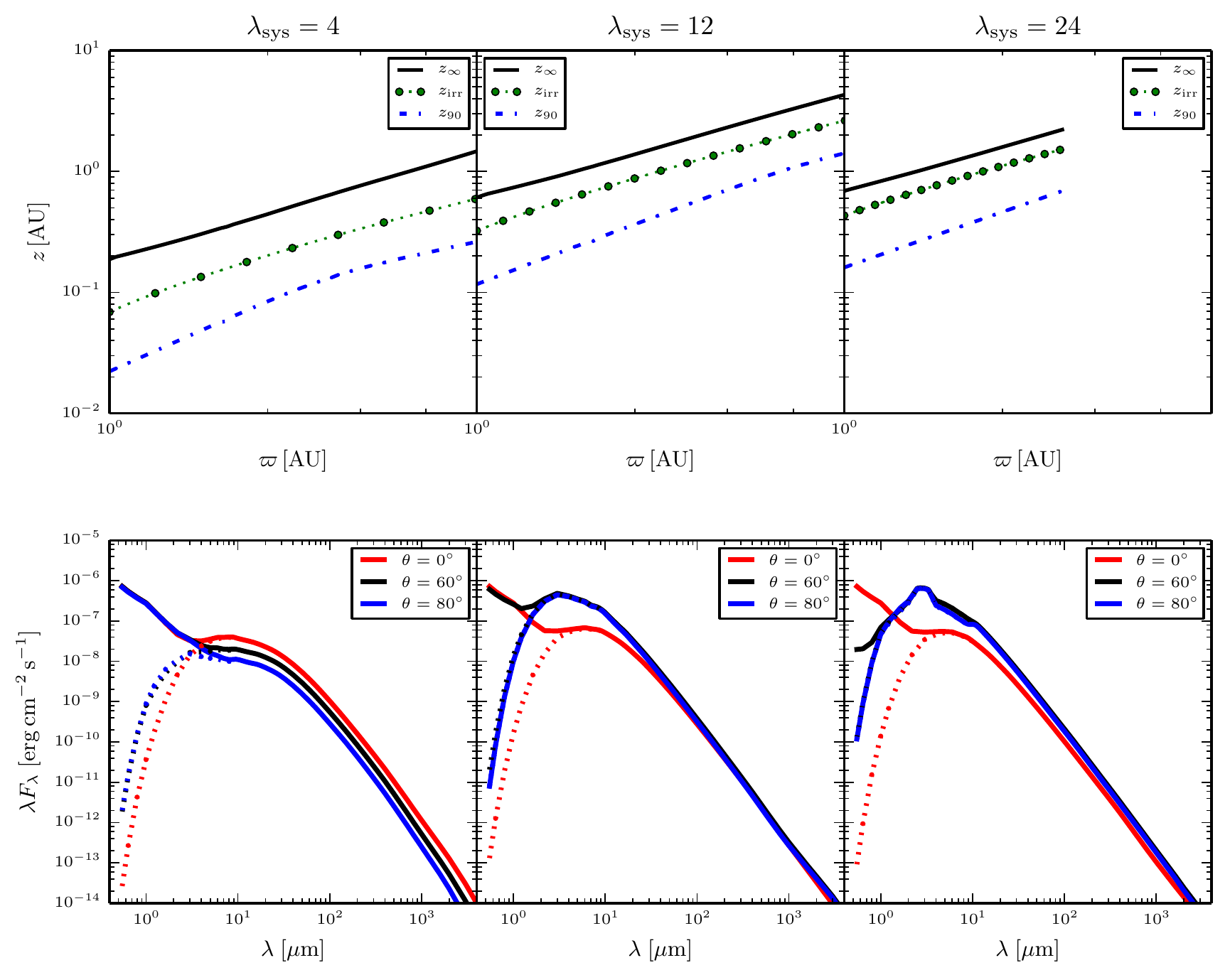}
\caption{
FU Ori disks with different mass-to-flux ratios, $\lambda_{\rm sys}=4, \, 12, \, 24$, that label each column.
The upper panels show the  the different disk surfaces.
 The lower panels show the SED of the star plus disk system at different inclination angles.
 Same description as Figure 2.
}
\label{fig:FUOriSED}
\end{figure}

\begin{figure} 
\centering
\includegraphics[angle=0,width=1.\textwidth]{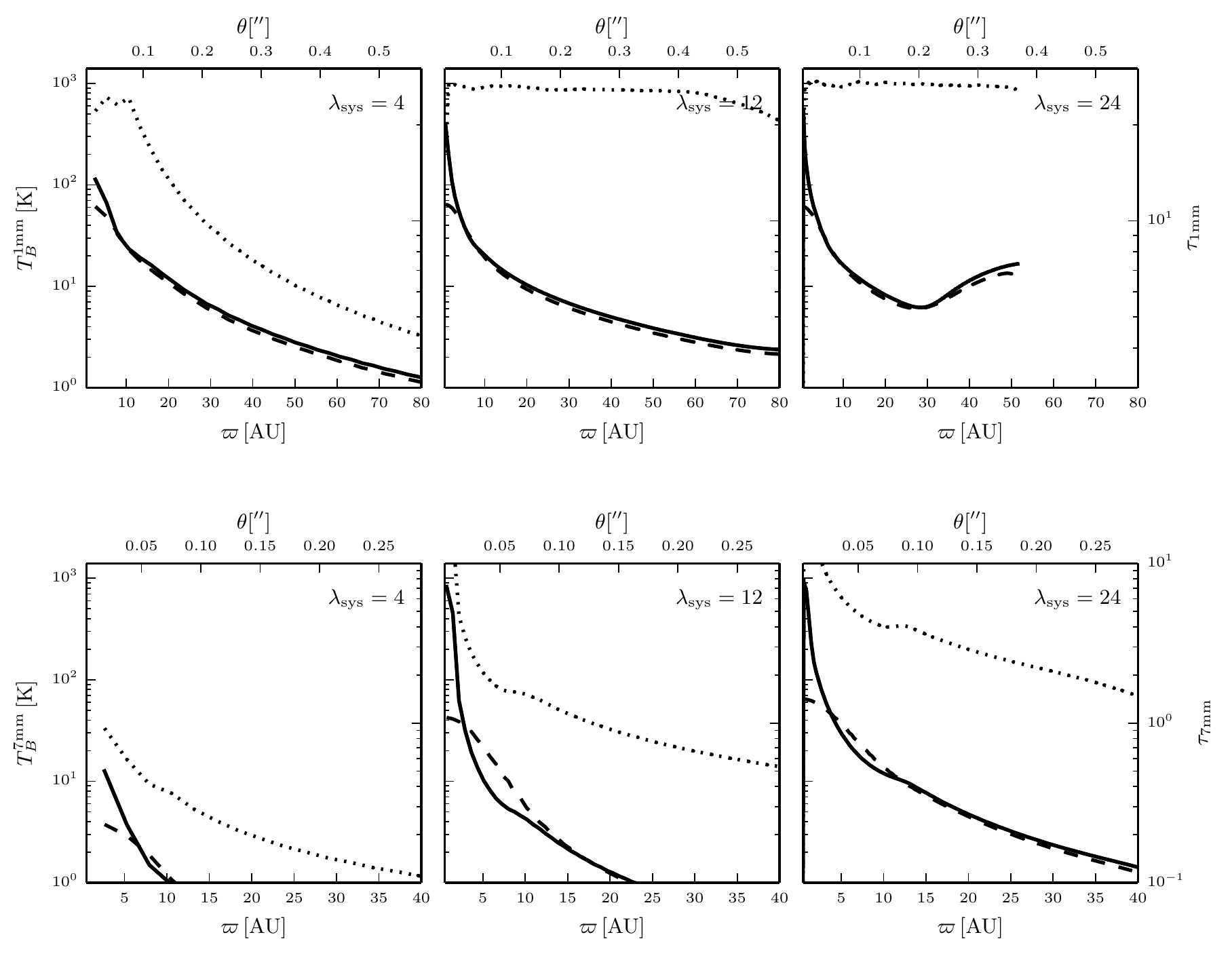}
\caption{
Averaged antenna temperature $T_B$ and optical depth $\tau_\lambda$ profiles of the LMP disk models with different 
mass-to-flux ratios, $\lambda_{\rm sys}= 4, \, 12, \, 24$,  as a function of
distance to the central star. The 1 mm and 7 mm profiles are shown in the upper and lower panels, respectively.
The disk models have an inclination of $\theta= 60^\circ$. 
The solid lines in each panel correspond to the antenna temperature profiles.  
The dashed lines correspond to the antenna temperature profiles 
convolved with the ALMA beam at 1 mm, $\theta_{ALMA} = 0.034 \arcsec$, and the VLA beam at 7 mm, $\theta_{7 \, mm} = 0.043 \arcsec$,
respectively. 
The dotted lines correspond to the optical depth, the values are shown in the left axis in each panel.
The upper axes gives the distance to the center in arc seconds, assuming a distance to the source of 140 pc. }
\label{fig:LMPprofiles}
\end{figure}

\begin{figure} 
\centering
\includegraphics[angle=0,width=1.\textwidth]{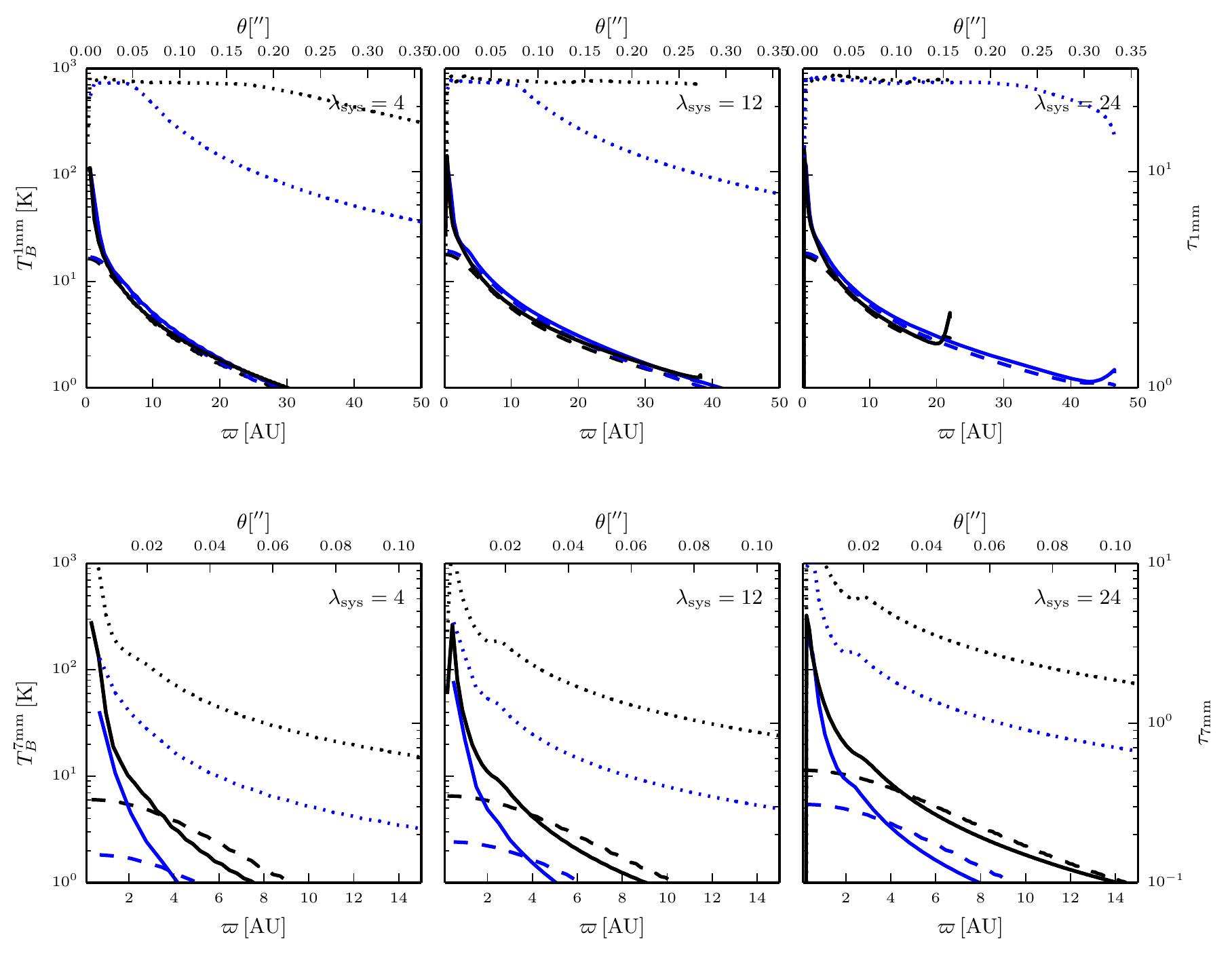}
\caption{Antenna temperature $T_B$ and optical depth $\tau_\lambda$ at 1 mm and 7 mm of T Tauri disks
 with different mass-to-flux ratios,  $\lambda_{\rm sys}= 4, \, 12, \, 24$,  as a function of distance to the central star.
The description of the panels and the lines is the same as in Figure {\ref{fig:LMPprofiles}}.  
The black color lines correspond to models with a viscosity coefficient $D=10^{-2.5}$, and 
the blue color lines correspond to a viscosity coefficient $D=0.01$.} 
\label{fig:TTauriprofiles}
\end{figure}

\begin{figure} 
\centering
\includegraphics[angle=0,width=1.\textwidth]{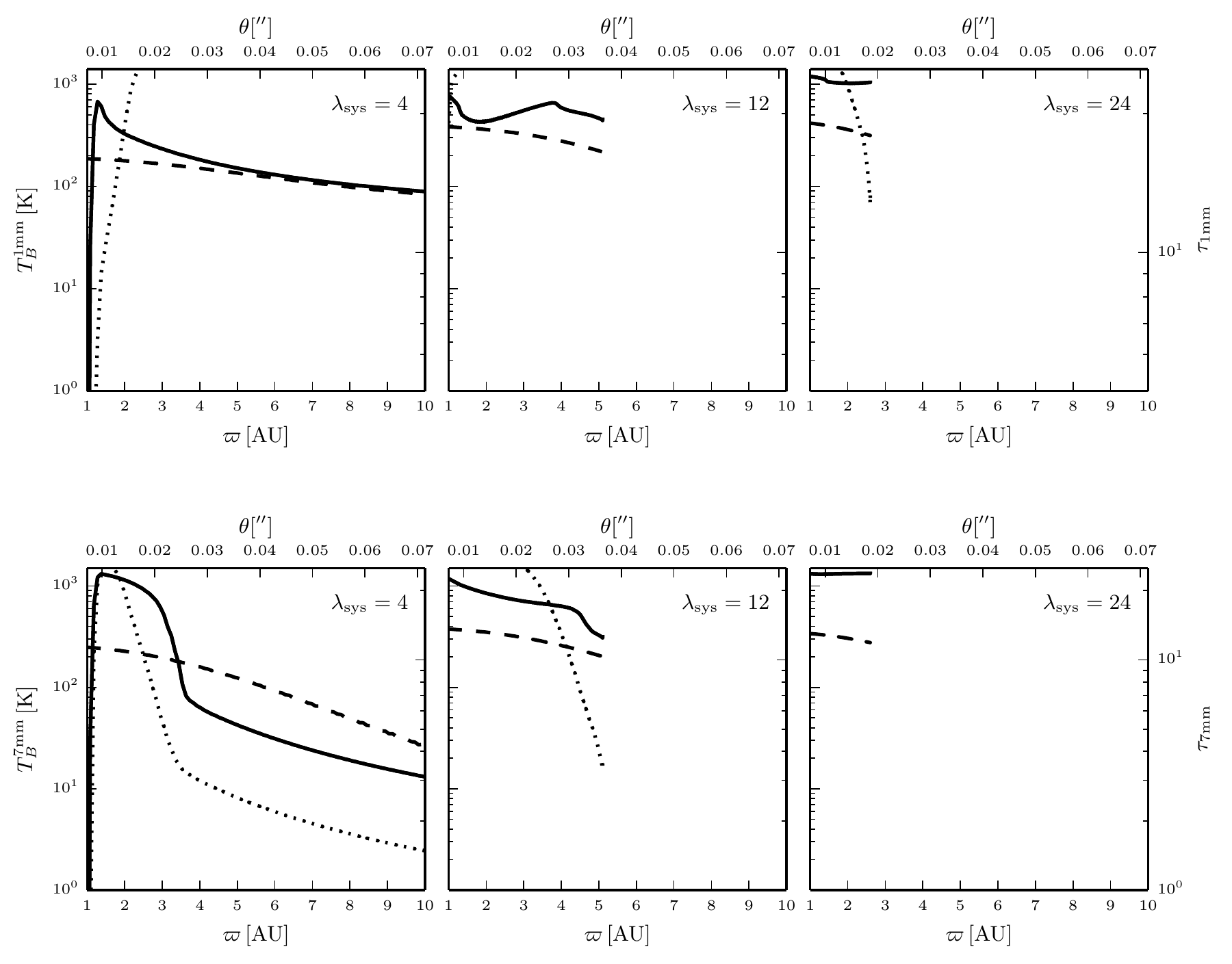}
\caption{Antenna temperature $T_B$ and optical depth $\tau_\lambda$ at 1 mm and 7 mm of FU Ori disks
with different mass-to-flux ratios,  $\lambda_{\rm sys}= 4, \, 12, \, 24$,  as a function of distance to the central star. 
The description of the panels and the lines is the same as in Figure {\ref{fig:LMPprofiles}}. 
 The upper middle panel and the lower right hand panel  do not show the
optical depth profile because, for numerical convenience, the
integration of the optical depth ends at the value 25.}
\label{fig:FUOriprofiles}
\end{figure}

\begin{figure} 
\centering
\includegraphics[angle=0,width=1.\textwidth]{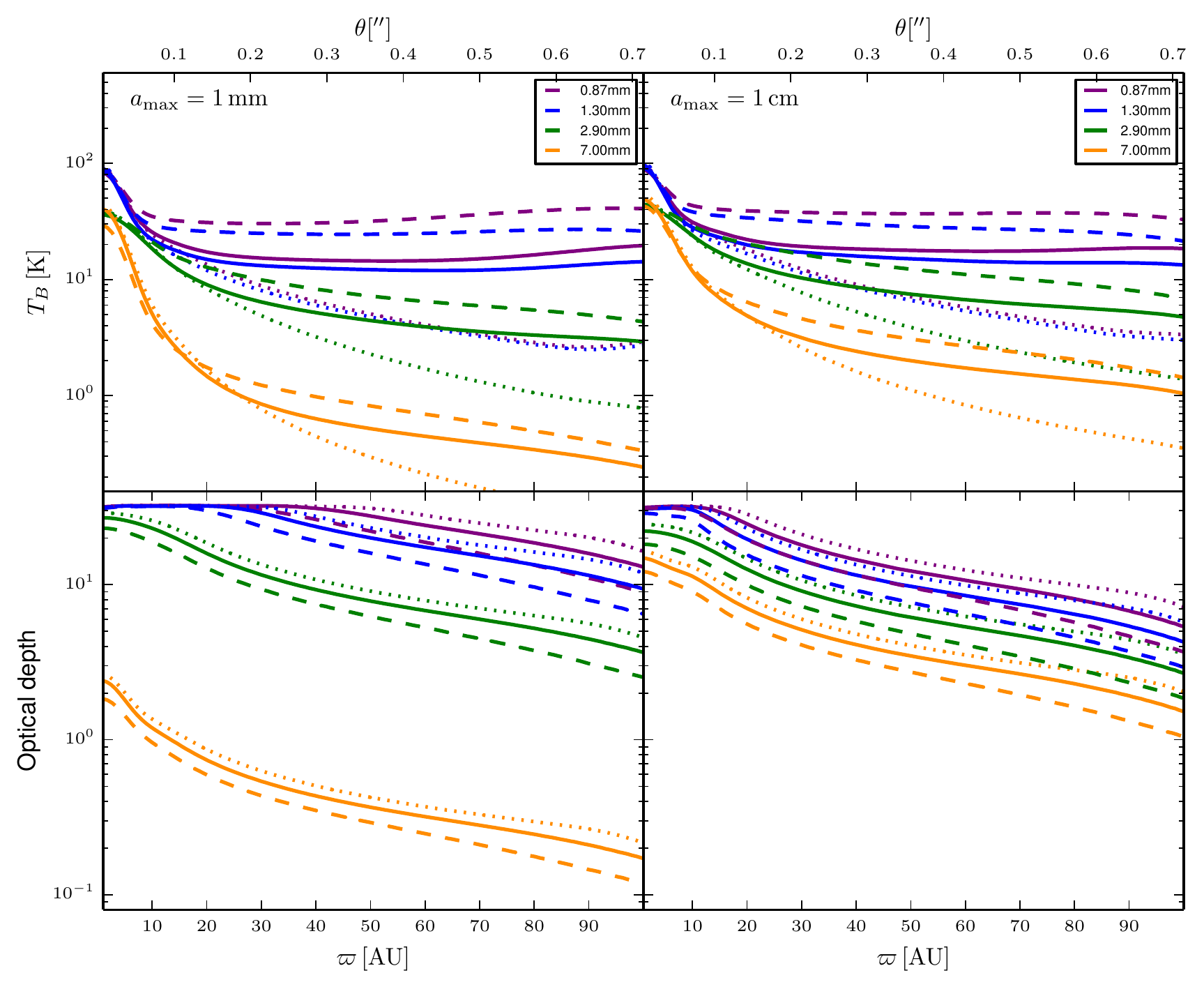}
\caption{
Antenna temperature $T_B$ and optical depth  
$\tau_\lambda$ at 0.87, 1.3, 2.9, and  7 mm of  
LMP disks as a function of distance to the central star. 
The radial profiles have been convolved with the  ALMA beams:  $\theta_{\rm 0.87\, mm} = 0.034 \arcsec$,   $\theta_{\rm 1.3\, mm} = 0.029 \arcsec$,  $\theta_{\rm 2.9\,mm} = 0.066 \arcsec$,  and the VLA beam at 7 mm, $\theta_{\rm 7 \, mm} = 0.043 \arcsec$. 
The color code is shown in the boxes in the upper right corners. This color code is similar to the one used in Figure 3 of
Carrasco-Gonz\'alez et al. (2016) that plots the observed ALMA and VLA antenna temperature profiles of HL Tau.
The disks have a  mass-to-flux ratio $\lambda_{\rm sys}=24$ . The disk parameters are shown in Table 4.
The upper panels show $T_B$ and the lower panels show $\tau_\lambda$ for each wavelength. 
The left panels correspond to models with a dust grain distribution with $a_{\rm max} = 1$ mm: 
 Model I (dotted lines) without envelope heating  ($T_e = 0$ K); Model II (solid lines) with $T_e = 50$ K;
Model III (dashed lines) with  $T_e = 100$ K.
The right panels correspond to models with  $a_{\rm max} = 1$ cm: 
 Model IV (dotted lines) without envelope heating  ($T_e = 0$ K); Model V (solid lines) with $T_e = 50$ K;
Model VI (dashed lines) with  $T_e = 100$ K. 
}
\label{fig:HLTauprofiles}
\end{figure}

 \end{document}